\documentclass[aps,reprint,superscriptaddress]{revtex4-1}

\usepackage{graphicx}
\usepackage{physics}
\usepackage{braket}
\usepackage{hyperref}
\usepackage{amsfonts}
\usepackage{amssymb}
\usepackage{xcolor}
\usepackage{enumitem}
\usepackage{soul}

\graphicspath{{"../figures/"}}

\usepackage[normalem]{ulem}

\begin{document}

\title{
Quantum-amplified global-phase spectroscopy on an optical clock transition}
\author{Leon Zaporski\textsuperscript{1,*}}
\author{Qi Liu\textsuperscript{1,*}}
\author{Gustavo Velez\textsuperscript{1,*}}
\author{Matthew Radzihovsky\textsuperscript{1,*}}
\author{Zeyang Li\textsuperscript{1,2}}
\author{Simone Colombo\textsuperscript{1,3}}
\author{Edwin Pedrozo-Pe\~nafiel\textsuperscript{1,4}}
\author{Vladan Vuleti\'c\textsuperscript{1,$\dagger$}}

\noaffiliation 

\affiliation{MIT-Harvard Center for Ultracold Atoms and Research Laboratory of Electronics, Massachusetts Institute of Technology, Cambridge, Massachusetts 02139, USA }
\affiliation{Department of Applied Physics, Stanford University, Stanford, CA}
\affiliation{Department of Physics, University of Connecticut, 196A Auditorium Road, Unit 3046, Storrs, Connecticut 06269-3046, USA}
\affiliation{Department of Physics, University of Florida, Gainesville, FL 32611, USA
\\ \ \\
\textsuperscript{*}\,These authors contributed equally.\\
\textsuperscript{$\dagger$}\,Correspondence to: vuletic@mit.edu.
\\ \ \\
}

\begin{abstract}
Optical lattice clocks (OLCs) are at the forefront of precision metrology \cite{Ludlow15,Ushijima2015,Oelker2019,Schioppo2017,Pan2024,Robinson2024}, operating near a standard quantum limit (SQL) set by quantum noise \cite{Schioppo2017,Zheng2022}. Harnessing quantum entanglement offers a promising route to surpass this limit \cite{Pezze18,Monz11,Pogorelov21,Leibfried2005,Cao2024,Kitagawa93,Yurke86,Wineland94,Bollinger96}, yet there remain practical roadblocks concerning scalability and measurement resolution requirements \cite{Davis16,Frowis16}. Here, we adapt the holonomic-quantum-gate concept~\cite{Zhou2017,SJOQVIST201665} to develop a novel Rabi-type ``global-phase spectroscopy'' (GPS) that utilizes the detuning-sensitive global Aharanov-Anandan phase~\cite{Aharonov87}. With this approach, we are able to demonstrate quantum-amplified time-reversal spectroscopy on an optical clock transition that achieves directly measured $2.4(7)$~dB metrological gain, and $4.0(8)$~dB improvement in laser noise sensitivity beyond the SQL.
To this end, we introduce rotary echo to protect the dynamics from inhomogeneities in light-atom coupling and implement a laser-noise-canceling differential measurement through symmetric phase encoding in two nuclear spin states. Our technique is not limited by measurement resolution, scales easily owing to the global nature of entangling interaction, and exhibits high resilience to typical experimental imperfections. 
We expect it to be broadly applicable to next-generation atomic clocks and other quantum sensors approaching the fundamental quantum precision limits \cite{Kaubruegger21,Thurtell2024,Qi2025,Marciniak2022}. 

\end{abstract}
\maketitle

The progress of science is largely determined by the level of measurement sensitivity to increasingly weaker signals. Among humanity's most precise sensors, optical lattice clocks (OLCs) have reached unprecedented fractional frequency instability and inaccuracy at the $10^{-18}$ level \cite{Ludlow15,Ushijima2015,Oelker2019,Schioppo2017,Pan2024,Robinson2024}. Besides serving as precise time references, OLCs have paradigmatic applications in advancing relativistic geodesy \cite{Grotti2018,Mehlstaubler2018,Bothwell2022}, detecting variations of fundamental constants~\cite{Safronova18} and gravitational waves~\cite{Kolkowitz16}, testing Lorentz invariance~\cite{Sanner2019}, and searching for dark matter~\cite{Wcislo18}. 

\begin{figure*}[htp]
    \centering
    \includegraphics[width=0.9\textwidth]{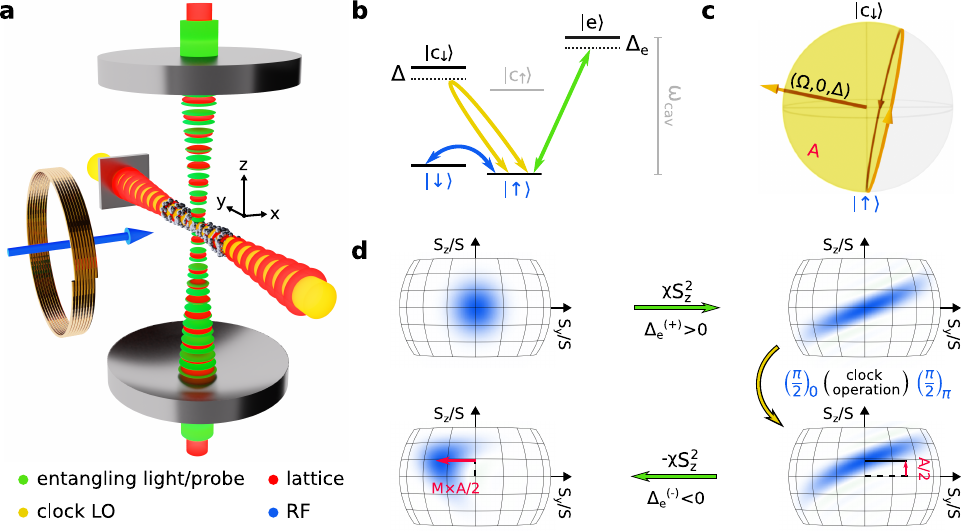}
    \caption[width=1\textwidth]{\textbf{Experimental setup for entangled time-reversal global-phase spectroscopy (GPS)} \textbf{a,} Laser-cooled atoms are confined in a 2D optical lattice (red) inside a high-finesse optical cavity. A transverse lattice along the $y$ axis is created via retro-reflection off the (square) mirror. Squeezing and probing light (green) is sent through the cavity along the $z$ axis, while the clock laser (yellow) is aligned with the transverse lattice. 
    The blue arrow represents the RF field used for rotations of the nuclear Zeeman ground states $\{ \ket{\uparrow}, \ket{\downarrow} \}$. The objects in the picture are rendered not to scale. \textbf{b,} Simplified energy level diagram with states $\ket{\downarrow}\equiv \ket{{}^1\text{S}_0, m_I=-\frac{1}{2}}$, $\ket{\uparrow}\equiv \ket{{}^1\text{S}_0, m_I=+\frac{1}{2}}$, $\ket{\text{c}_\downarrow}\equiv \ket{{}^3\text{P}_0, m_I=-\frac{1}{2}}$, $\ket{\text{c}_\uparrow}\equiv \ket{{}^3\text{P}_0, m_I=+\frac{1}{2}}$, $\ket{\text{e}}\equiv \ket{{}^3\text{P}_1, m_F=+\frac{3}{2}}$, quantized along the $z$ axis from panel \textbf{a}. The arrows represent the three control fields from panel \textbf{a}. The cavity mode (vertical gray line) is tuned on resonance with the $\ket{\uparrow}\rightarrow\ket{e}$ transition. 
    \textbf{c,} In GPS, the clock laser drives the optical qubit state along the closed trajectory (yellow curve) on the unit-radius Bloch sphere, encoding a geometric Aharonov-Anandan phase, which equals half of the enclosed area, $A/2$, after a single cyclic evolution. The area $A$ depends on clock laser oscillator (LO) detuning, providing a method to measure LO frequency while operating nominally on atomic resonance. \textbf{d,} Illustration of quantum amplification based on time-reversal. The initial coherent spin state (CSS) is squeezed by a green laser pulse and rotated by $\pi/2$ by an RF pulse before the clock LO pulse is applied to induce a shift of $A/2$ that is mapped back onto the $S_z$ axis by another $\pi/2$ RF pulse. Unsqueezing $-\chi S_z^2$ subsequently amplifies the signal to $M\times A/2$ along the $S_y$ axis (with $M$ constant and independent of $A$ in the vicinity of $A=0$ and $A=2\pi$). The blue shading on the generalized Bloch spheres illustrates the Wigner quasi-probability distributions in the ground state manifold, $\{\ket{\uparrow},\ket{\downarrow}\}^{\otimes N}$.
    } 
    \label{fig:figure_1}
\end{figure*}

The short-term stability of optical clocks is guaranteed by locking clock lasers to ultra-stable cavity references, while the cancelation of long-term laser frequency drift entails closed-loop stabilization to the atomic clock transition. Here, the relative phase between laser and atoms due to a detuning from atomic resonance is converted into a population imbalance that can be directly measured.
The accuracy and stability of the clock are  conditioned by the precision with which this phase can be estimated. 
For an ensemble of $N$ independent atoms, the single-shot uncertainty in phase estimation is fundamentally limited by quantum projection noise, $\delta \phi=1/\sqrt{N}$, known as the standard quantum limit (SQL) \cite{Pezze18}. State-of-the-art OLCs have reached this limit after suppressing the Dick noise, either by shortening dead time~\cite{Schioppo2017} or using synchronous differential comparisons in multiplexed clocks~\cite{Zheng2022}. By harnessing multi-particle entanglement this precision cap can be further lowered to the Heisenberg limit of $\delta \phi = 1/N$ -- the ultimate sensitivity allowed by quantum mechanics \cite{Pezze18}. Highly entangled resources, such as Greenberger–Horne–Zeilinger (GHZ) states, have been prepared and have approached this limit in small-scale systems~\cite{zhushiyao2019,Lukin2019,Finkelstein2024}, and were very recently leveraged in OLCs to achieve sub-SQL frequency instability~\cite{Cao2024}. However, GHZ states are difficult to prepare and maintain for large atom numbers.

Squeezed states \cite{Kitagawa93,Yurke86,Wineland94,Bollinger96} are another widely explored class of entangled states. They feature wider dynamic range and higher tolerance to decoherence and loss compared to GHZ states, at the expense of intermediate levels of improvement beyond the SQL. Thus far, spin squeezing has been generated in microwave-coupled manifolds~\cite{Gross2010,Riedel2010,Hamley2012,Hosten2016,Bao2020}, in momentum space~\cite{Klempt2021,Greve2022} and on optical transitions~\cite{Pedrozo-Penafiel2020,Marciniak2022}, where it survived well beyond tens of milliseconds~\cite{Pedrozo-Penafiel2020}. Spin squeezing has applications in atomic magnetometers~\cite{Sewell2012,Muessel14}, gravimeters~\cite{Klempt2025} and clocks~\cite{Kruse16,Pedrozo-Penafiel2020,Appel09}. Spin squeezed states can either be detected directly, which requires measurement resolution below the quantum projection noise limit, or by further manipulation of the entangled state after phase accumulation, also known as interaction-based readout~\cite{Haines2017_2}. The latter includes quantum phase amplification techniques~\cite{Hosten16mag,Linnemann16,Gilmore21,Colombo2022,Liu2022}, typically based on time-reversed squeezing dynamics, which can be used to approach the Heisenberg limit without single-particle resolution~\cite{Davis16,Frowis16,Haines2017}. Extensions of the interaction-based readout emerged from variational optimization of Ramsey interferometry~\cite{Kaubruegger21,Thurtell2024,Qi2025} to balance local sensitivity and dynamic range while minimizing the frequency Allan deviation, and were recently demonstrated on an optical transition in trapped ions~\cite{Marciniak2022}. Despite its potential significance for OLCs, the interaction-based readout on an optical clock transition in a neutral-atom system has not been demonstrated to date.

Here, we report the first experimental quantum amplification of an optical clock phase in a neutral atom ensemble and demonstrate spectroscopic precision enhancement of $2.4(7)$ dB beyond the SQL. This is achieved through cavity-mediated one-axis twisting (OAT)~\cite{Schleier-Smith2010a} which generates squeezing and unsqueezing dynamics in the nuclear ground states of laser-cooled ${}^{171}$Yb atoms~\cite{Braverman2019}. To mitigate fast phase diffusion on the optical transition due to high-frequency noise of the local oscillator (LO) laser~\cite{Bishof13,supp}, we develop a new method that replaces conventional Ramsey spectroscopy with a novel Rabi-type ``global-phase spectroscopy'' (GPS) that relies on driven cyclic evolution. In this approach, the driven optical qubit acquires a global Aharonov-Anandan phase~\cite{Aharonov87}, which realizes a detuning-sensitive holonomic quantum phase gate \cite{Zhou2017,SJOQVIST201665} between the ground states.
This new GPS method allows us to extend for the first time entanglement enhancement techniques to Rabi-type spectroscopy. 
While it retains the same dc sensitivity to the LO laser noise as conventional Rabi spectroscopy, the GPS features reduced sensitivity to the high-frequency phase (frequency) noise, following ${\propto}f^{-4}$ (${\propto}f^{-6}$) scaling. This is substantially lower than that of Rabi spectroscopy, ${\propto}f^{-2}$ (${\propto}f^{-4}$), and Ramsey spectroscopy, ${\propto}f^{0}$ (${\propto}f^{-2}$)~\cite{supp}.
Furthermore, while conventional Rabi spectroscopy measures population imbalance, which necessitates side-of-fringe operation, GPS measures phase and exhibits maximal sensitivity on resonance - an optimal condition for feedback. This fact allows us to also integrate GPS with a resonant rotary echo \cite{Solomon59,yan2025}, which refocuses the inhomogeneities in light-atom coupling \cite{Blatt09}, and facilitates implementation of composite pulse sequences.

To characterize the metrological gain available to stabilize a noisy LO laser, we leverage the multi-level structure of ${}^{171}$Yb by performing a differential measurement on two clock transitions in a single ensemble.
The differential phase imprinted on a squeezed probe state is amplified by a time-reversal protocol~\cite{Davis16,Colombo2022,Hosten16mag}, leading to a metrological gain of $2.4(7)$ dB below the SQL ($4.0(8)$ dB when subtracting the residual laser noise), the first such demonstration in a scalable neutral atom system with global entangling interactions.

Our experiments are performed with an ensemble of $N = 2.2(4)\times 10^2$ laser-cooled ${}^{171}$Yb atoms that are trapped in a two-dimensional optical lattice inside a high-finesse optical cavity (Fig.~\ref{fig:figure_1}\textbf{a}). We initialize the atoms in the $\ket{\uparrow}\equiv \ket{{}^1\text{S}_0,m_I=+\frac{1}{2}}$ state, and rotate them into a coherent superposition of $\ket{\uparrow}$ and $\ket{\downarrow}\equiv\ket{{}^1\text{S}_0,m_I=-\frac{1}{2}}$ states with a resonant RF driving field. Optical phase encoding involves back-and-forth transfer of the $\ket{\uparrow}$ state or $\ket{\downarrow}$ state to one of the  $\{\ket{\text{c}_{\uparrow}}\equiv\ket{{}^3\text{P}_0,m_I=+\frac{1}{2}},\ket{\text{c}_{\downarrow}}\equiv\ket{{}^3\text{P}_0,m_I=-\frac{1}{2}}\}$ clock states (Fig.~\ref{fig:figure_1}\textbf{b}), and is achieved with a clock laser referenced to a commercial rack-mounted ultra-low-expansion (ULE) cavity. In this process, a near-resonant Rabi pulse drives the optical-qubit state around the $(\Omega,0,\Delta)$ axis to traverse a closed trajectory on the Bloch sphere (Fig.~\ref{fig:figure_1}\textbf{c}), where $\Omega$ and $\Delta$ stand for the resonant Rabi frequency and the laser detuning, respectively. The state of an atom evolves according to
\begin{equation}\label{eq:psi_t}
    \ket{\psi(\tau)}=\tfrac{1}{\sqrt{2}}\left[\ket{\downarrow}+e^{-i\frac{\Delta}{2}\tau}\left(a_\uparrow(\tau)\ket{\uparrow}+a_{\text{c}_\downarrow}(\tau)\ket{\text{c}_\downarrow}\right) \right],
\end{equation}
and by the end of a single cyclic evolution ($\tau \times\sqrt{\Omega^2+\Delta^2}=2\pi$), the coupled ground state $\ket{\uparrow}$ recovers its initial population ($a_{c_\downarrow} \approx 0$), acquiring a detuning-dependent geometric phase
\begin{equation}\label{eqn:aa_phase}
    \phi = -\pi- \frac{\Delta}{2} \tau = -\pi \left(1 + \frac{\Delta}{\sqrt{\Omega^2+\Delta^2}}  \right),
\end{equation}
known as the Aharonov-Anandan phase \cite{Aharonov87}, whose value corresponds to half the area enclosed by the trajectory (see Methods).
The global phase of the optical qubit $\{ \ket{\uparrow}, \ket{\text{c}_\downarrow} \}$ thus translates to the relative phase between the $\ket{\uparrow}$ and $\ket{\downarrow}$ states, and is mapped with an RF rotation into a population difference between $\ket{\uparrow}$ and $\ket{\downarrow}$, that is measured via the cavity (see Methods).
Overall, the clock laser detuning from the atomic transition $\Delta$ is thus measured via the accumulated global phase $\phi$.

\begin{figure}
    \centering
    \includegraphics[width=0.48\textwidth]{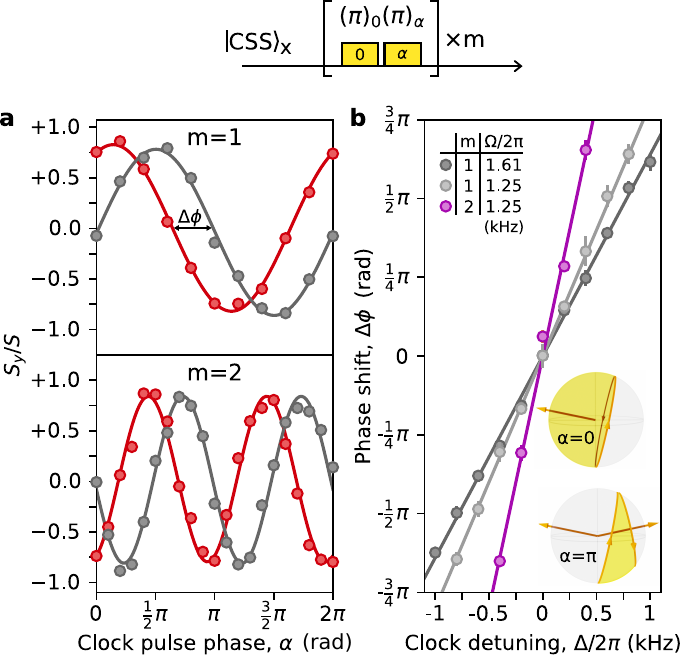}
    \caption{\textbf{Optical global phase encoding and readout.} \textbf{a,} A geometric phase is imprinted onto the $\ket{\uparrow}$ state using strings of $m$ pairs of clock $\pi$-pulses with alternating phases, $0$ and $\alpha$, and is mapped onto the normalized collective spin $S_y/S$. The gray and red points depict $S_y/S$ as a function of $\alpha$ for resonant and off-resonant optical driving, respectively.
    Here, $\Omega/(2\pi)= 4.5$~kHz for all the datasets, as well as $\Delta/(2\pi)= 1$~kHz for $m=1$ and $\Delta/(2\pi)= 1.2$ kHz for $m=2$. 
    \textbf{b,} Dependence of the phase shift, $\Delta \phi$, on the clock laser detuning for different Rabi frequencies and numbers $m$ of pulse pairs. The lines represent the theoretical prediction $\Delta\phi =\pi m \Delta/\Omega$ for $\Delta/\Omega \ll 1$.
    Throughout the manuscript, the labels above the pulses in the pulse diagrams, of the form $(\theta)_\varphi$, specify the pulse amplitude (in terms of the resonant rotation angle, $\theta$, expressed in radians), and phase ($\varphi$, expressed in radians).
    \textbf{Inset,} Trajectories traversed on the Bloch spheres for off-resonant Rabi pulses for $\alpha=0$ and $\alpha=\pi$. Note that for nonzero $\alpha$, the encoded geometric phase equals half the area enclosed by the unclosed trajectory and the shortest geodesic connecting the initial and final points~\cite{zhou2024geometric}. Error bars indicate one standard deviation in this and the following figures.
    }
    \label{fig:figure_2}
\end{figure}

The above sequence can be performed with unentangled coherent spin states (CSSs) or with spin squeezed states. To generate spin squeezing, we adopt the cavity feedback method~\cite{Braverman2019}. In essence, we tune the cavity mode to resonance with the $\ket{\uparrow}\to\ket{\text{e}}\equiv \ket{{}^3\text{P}_1,m_F=+\frac{3}{2}}$ transition, and send an off-resonant entangling light pulse through the cavity. The atoms in the $\ket{\uparrow}$ state dispersively shift the cavity resonance, which in turn alters the intra-cavity photon number. This leads to a spin-imbalance-dependent light shift captured by the OAT Hamiltonian~\cite{Braverman2019}:

\begin{equation}
\hat{H}=\chi\hat{S}_z^2.
\label{Eq:Hamiltonian}
\end{equation} 
Here, $\hat{S}_z$ is a collective spin operator, summed over the spins of all the atoms, while $\chi$ denotes the squeezing strength. This all-to-all interaction shears the noise distribution of the initial CSS into that of a spin-squeezed state, as shown by the Wigner quasi-probability distributions in Fig. \ref{fig:figure_1}d. To realize a process that is tolerant to detection noise and can work with oversqueezed states, we implement effective time-reversal (unsqueezing) by flipping the sign of the entangling light detuning and of the Hamiltonian Eq. \ref{Eq:Hamiltonian}~\cite{Colombo2022}.  The squeezing-unsqueezing sequence converts a small optical phase shift $\Delta \phi$ into a shift along $S_z$ that is amplified into a larger shift along the orthogonal quadrature ($S_y$), thereby achieving phase sensitivity below the SQL. 

We start by benchmarking our new GPS method on the optical clock transition. To this end, we initialize the atoms in a CSS polarized along the $S_x$-axis of the ground-state Bloch sphere, $\{ \ket{\uparrow}, \ket{\downarrow} \}$, and drive a cyclic evolution between $\ket{\uparrow}$ and $\ket{\text{c}_{\downarrow}}$ using two consecutive $\pi$-rotations with relative phase $\alpha$. In the near-resonant case ($\Delta\ll\Omega$), the $\ket{\uparrow}$ state acquires a geometric phase $\Delta \phi$ given by $\Delta \phi= \alpha+\pi(1+\Delta/\Omega)$. (For $\Delta=0$ and $\alpha=0$ this is just the minus sign that a spin-$\frac{1}{2}$ consisting of states $\{ \ket{\uparrow}, \ket{ \text{c}_{\downarrow}} \}$ acquires upon a $2\pi$ Rabi rotation.)
The phase $\Delta \phi$ is reflected in the measured normalized spin operator $S_y/S$, which oscillates as $\alpha$ is varied from $0$ to $2\pi$. The top two datasets in Fig. \ref{fig:figure_2}a illustrate the dependence of $S_y/S$ on $\alpha$ for both resonant (gray points) and off-resonant (red points) driving, which are related by a phase offset $\Delta\phi$, from which the laser detuning can be inferred.
As we vary the laser detuning, we observe a linear relationship of the fitted phase shifts shown in Fig. \ref{fig:figure_2}b, in good agreement with the theoretical expectation (solid lines). This linear frequency discriminant is a fundamental ingredient of locking the clock laser to the atoms, and can be made steeper by extending the interrogation time either through lowering the Rabi frequency or driving a repeated cyclic evolution with $m>1$. 
Extending the sequence to $m$ cycles multiplies the resulting phase and detuning sensitivity by a factor $m$, as showcased by the bottom two datasets in Fig. \ref{fig:figure_2}a (for $m=2$). Our theoretical analysis presented in the Supplement \cite{supp} reveals that a single-cycle sequence is preferred for dc measurements, while a multi-cycle sequence can enhance the ac signal sensitivity at particular frequencies.

Single-shot estimation of laser frequency requires maximizing the slope $\lvert\text{d}S_y/\text{d}\Delta\rvert$ around $\Delta=0$, which is reached at $\alpha\in\{0,\pi\}$ and $\alpha\in\{0,\frac{\pi}{2},\pi,\frac{3\pi}{2} \}$ for sequences with $m=1$ and $m=2$, respectively.
While the $\alpha=0$ case is simply a Rabi oscillation, the $\alpha=\pi$ realizes a rotary echo \cite{Solomon59}, conventionally applied in solid-state spin systems for extending the coherence time of the drive.
In our system, rotary echo ensures a better $\ket{\uparrow}$ state population recovery in the presence of inhomogeneous broadening in the atomic cloud, maintaining high signal contrast and enabling near-perfect time-reversal. 

%

%

\begin{figure}
    \centering
    \includegraphics[width=0.5\textwidth]{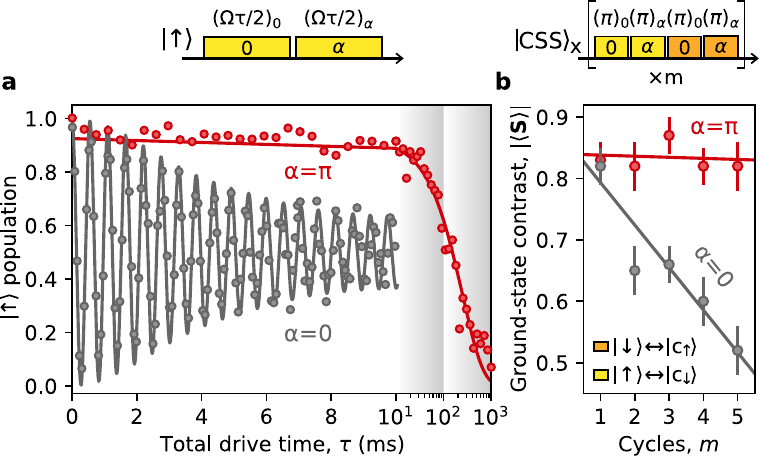}
    \caption{\textbf{Rabi and rotary echo sequences on the clock transition}. \textbf{a,} The gray (red) points show the $\ket{\uparrow}$ population oscillation under the resonant Rabi, $\alpha=0$, (rotary echo, $\alpha=\pi$) sequence with $\Omega/(2\pi)= 1.8$~kHz. The solid lines represent the models of population dynamics factoring the finite temperature effects and slight beam misalignment in agreement with ref. \cite{Blatt09}, as well as the exponential decay of the clock state over the lifetime of $0.25(2)$ seconds. 
   \textbf{b,} The ground-state contrast measured after $m$ cycles of resonant, symmetric phase encoding with $\Omega/(2\pi)= 3.8$ kHz. The red (gray) points correspond to the rotary echo sequence (Rabi sequence). The solid lines represent linear fits. The rotary-echo sequence ($\alpha=\pi$) retains high contrast in spite of inhomogeneously broadened atom-light coupling. Faster contrast decay of the Rabi sequence (gray points), compared to the panel a, is likely due to an increase in misalignment angle (from 3.5 mrad to 7 mrad). 
   } 
    \label{fig:figure_3}
\end{figure}

The effectiveness of the rotary echo sequence in our setup is explored in Fig. \ref{fig:figure_3}. We first obtain a simple resonant Rabi sequence benchmark by initializing the system in the $\ket{\uparrow}$ state and driving the atoms through the $\ket{\uparrow}\rightarrow\ket{\text{c}_{\downarrow}} \rightarrow \ket{\uparrow}$ transition. The gray points in Fig. \ref{fig:figure_3}a show the optical Rabi oscillation at $\Omega/(2\pi) = 1.8$ kHz, decaying over the timescale of a few milliseconds. This damping can be attributed to the finite temperature of the atomic ensemble, which populates multiple motional states of the transverse lattice, leading to inhomogeneous coupling to the clock laser~\cite{Blatt09}.
The dephasing is remarkably suppressed when using rotary echo. As shown in Fig. \ref{fig:figure_3}a (red points), we observe a robust refocusing over hundreds of milliseconds, constrained by the lifetime of the clock state of $T=0.25(2)$s due to photon scattering from the trapping light. 

We now turn to a measurement of the atomic clock phase via GPS.
To measure the intrinsic atomic phase stability without being overwhelmed by LO laser noise, we apply our GPS method sequentially on both the $\ket{\uparrow}\to\ket{\text{c}_{\downarrow}}$ and $\ket{\downarrow}\to\ket{\text{c}_{\uparrow}}$ transitions, starting from a CSS in the $\{ \ket{\uparrow}, \ket{\downarrow} \}$ manifold polarized along the $S_x$ axis (see the pulse sequence above Fig. \ref{fig:figure_3}b). We repeat this over multiple cycles for both the Rabi ($\alpha=0$) and the rotary echo ($\alpha=\pi$) sequences and map the resulting ground-state Bloch vector lengths, $|\langle \mathbf{S}\rangle|$, to display contrast loss. As illustrated in Fig. \ref{fig:figure_3}b, the rotary echo maintains the contrast at the level of ${\sim}81\%$ for at least $m=5$ cycles (red points), proving negligible propagation of the inhomogeneities in light-atom coupling to the geometric phases.
In stark comparison, the Rabi sequence features fast contrast decay (gray points).
These results illustrate that rotary-echo GPS can be extended to multi-cycle composite sequences of long total probing times even in the presence of inhomogeneous broadening of the atom-LO coupling.

\begin{figure*}[t!]
    \centering
    \includegraphics[width=0.75\textwidth]{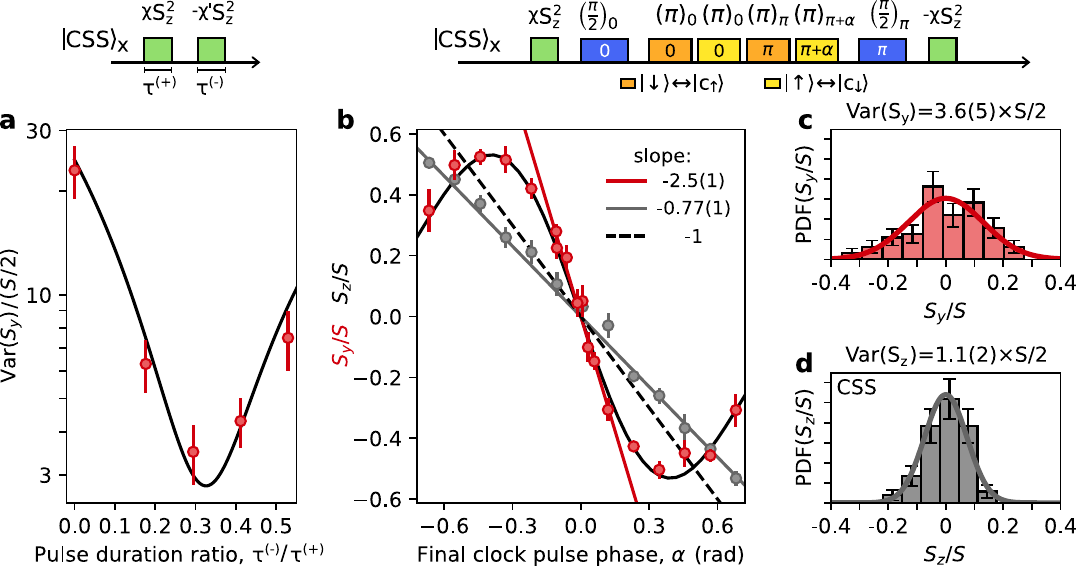}
    \caption{\textbf{Quantum amplification measurement of the optically encoded phase}. \textbf{a,} A pair of entangling pulses with durations $\tau^{(\pm)}$ generate the squeezing-unsqueezing dynamics in the ground state manifold. The red data points display normalized $S_y$-variances as a function of the pulse duration ratio $\tau^{(-)}/\tau^{(+)}$, with $\tau^{(+)}$ fixed to $8.5$ ms, in good agreement with the microscopic model indicated by the black line (see \cite{supp}).
    \textbf{b,} The red (gray) points represent the average values of $S_y/S$ ($S_z/S$) as a function of the final clock pulse phase in a time-reversal (CSS reference) differential measurement. The red and gray lines represent the linear fits of the slope at $\alpha=0$.
    The solid black line is a numerical simulation of the data, extending beyond the linear amplification regime. For the equivalent scenario of the clock detuning measurement, $\alpha$ would map to $\pi \times \Delta/\Omega$ (c.f. Fig. \ref{fig:figure_2}b).  
    \textbf{c,} Spin noise measured at $S_y/S=0$ after a time-reversal sequence. \textbf{d,} The CSS reference counterpart of dataset from panel \textbf{c}. The `PDF' in panels \textbf{c} and \textbf{d} stands for probability density function, estimated based on more than $100$ measurement outcomes summarized by the histograms. The color coding in the pulse diagrams is consistent with Fig. \ref{fig:figure_1}a-b (Green: entangling light, blue: RF, yellow/orange: clock laser).
   } 
    \label{fig:figure_4}
\end{figure*}

Equipped with a robust optical phase encoding and detection scheme, we proceed to demonstrate entanglement-enhanced operation of the GPS protocol below the SQL.
We set the entangling pulse detunings to $\Delta_{\text{e}}^{(+)}/(2\pi)= 8.33$ MHz and  $\Delta_{\text{e}}^{(-)}/(2\pi)=- 7.28$ MHz, respectively, for the squeezing and unsqueezing dynamics within the ground-state manifold, $\{\ket{\uparrow},\ket{\downarrow}\}$~\cite{Colombo2022}. These parameters were chosen to mitigate the effect of fluctuations in the total atom number, considering the contribution of the $\ket{\downarrow}\to \ket{{}^3\text{P}_1,m_F=+\tfrac{1}{2}}$ transition to the OAT dynamics (see \cite{supp}). 

We prepare a squeezed state whose $S_y$ variance is enlarged by a factor of $23(4)$ compared to the quantum projection noise of the CSS (Fig. \ref{fig:figure_4}a). 
During the effective time-reversed dynamics, this variance is reduced as we increase the duration $\tau^{(-)}$ of the second (unsqueezing) pulse. We obtain a minimal normalized variance of $3.5(7)$ for $\tau^{(-)}=0.29\times \tau^{(+)}$. This variance exceeds the ideal value of 1, mainly due to deleterious effect of the residual entanglement between the atomic spin and the light leaving the cavity~\cite{Braverman2019} (see \cite{supp}). 

Having completed the time-reversal calibration step, we lower the entangling light intensity by $40\%$ to the level for which the metrological gain is expected to peak. In order to reduce the impact of clock laser noise, we resort to a sequential differential phase measurement~\cite{Schioppo2017,Norcia19,Takamoto2011,Nicholson12,Robinson2024,Zheng2022} by implementing two identical rotary-echo GPS sequences on $\ket{\uparrow}\to\ket{\text{c}_{\downarrow}}$ and $\ket{\downarrow}\to\ket{\text{c}_{\uparrow}}$ transitions (see Methods). These two transitions are separated by a frequency difference of $\Delta_{\uparrow \downarrow} =2\pi \times 25.3$ kHz, allowing us to work with Rabi frequency of $\Omega=2\pi\times 4.55$ kHz with no transition cross-talk (see Methods). In order to minimize the interval between the pulses on the two transitions, we adopt an interleaved interrogation (see the pulse diagram above Fig. \ref{fig:figure_4}b-d).

We next characterize the signal enhancement due to entanglement. The red points in Fig. \ref{fig:figure_4}b compare the signal $S_y/S$ of the time reversal sequence with entangled states to the reference signal for the unentangled CSS (gray points) as we sweep the relative optical phase $\alpha$.
We observe an increased slope of $-2.5(1)$ near $\alpha=0$ in the time-reversal sequence, resulting in a $8.0(3)$ dB increase in the signal power, compared to the perfect unentangled sequence. 
We also note that time reversal amplifies the signal power by $10.2(4)$ dB compared to the CSS reference sequence without squeezing, which features a reduced slope of $0.77(1)$.

To verify that the observed quantum amplification offers phase sensitivity below the SQL, we also evaluate the spin noise in $S_y$ at the end of the sequence.
Figure \ref{fig:figure_4}\textbf{c} (\ref{fig:figure_4}\textbf{d}) shows the histogram of $S_y/S$ ($S_z/S$) measured in presence (absence) 
of squeezing-unsqueezing dynamics around $S_y/S=0$ ($S_z/S=0$). The CSS noise features a normalized variance of $1.1(2)$, consistent with the SQL (see Methods). In contrast, the normalized variance measured after the quantum amplification of $3.6(5)$ visibly exceeds its benchmark counterpart of $2.5(4)$. This is explained by the amplification of the residual laser noise in the sequential sequence at the level of $10\%$ of the SQL.

Finally, based on the measured slope and spin noise, we infer a directly measured metrological gain of $\mathcal{G}=(\partial_\phi S_y)^2/[2S(\Delta S_y)^2]=2.4(7)$ dB below the SQL and $4.0(8)$ dB with the laser noise subtracted. We stress that because our measurement realizes a comparison of two optical frequencies, the characterized level of performance of $4.0(8)$ dB directly translates to the improved measurement precision of the frequency of the clock LO laser relative to a single clock transition (see Fig. \ref{fig:figure_2}).
Our experiment constitutes another direct measurement of sub-SQL phase sensitivity on an optical clock transition, following the prior demonstration using spin squeezing mapping~\cite{Pedrozo-Penafiel2020} and Rydberg-generated GHZ states of $4$ atoms~\cite{Cao2024}.
However, we highlight that the method demonstrated here offers superior scalability to much larger systems, owing to the intrinsically all-to-all character of the entangling interaction. 
For the current experimental cycle time of $5$ s, we infer a frequency instability of $2.0\times10^{-13}/\sqrt{\tau}$ for the differential phase measurement. Promising routes to further improvement include implementing a GPS sequence with lower Rabi frequency to extend the phase interrogation time, reducing the dead time through reusing the atoms multiple times after cavity non-destructive measurement, and improving the LO performance itself.

%

In conclusion, we have demonstrated the first-ever quantum amplification of the phase encoded on an optical clock transition, and directly observed sensitivity below the SQL. We have achieved this while developing a new spectrosopic method that utilizes the concepts of holonomic quantum gates \cite{Zhou2017,SJOQVIST201665}, measuring the frequency-dependent global phase of the two-level clock system with reference to a third level. 
Importantly, this has allowed us to apply, for the first time, entanglement-enhanced metrology to Rabi-type spectroscopy. (Entanglement-enhanced measurement of the Rabi drive amplitude has been demonstrated in Ref. \cite{Bao2020}.) We have also identified the rotary echo in Rabi spectroscopy as a practical tool to suppress inhomogeneous broadening, enabling composite precision spectroscopy. 
In the future, we anticipate achieving further increased metrological gain by employing non-Gaussian probe states involving more atoms. Based on our previous work~\cite{Colombo2022}, the improvements should follow the Heisenberg scaling.
The phase amplification technique on the clock transition can also be extended to multiple ensembles, opening new avenues for quantum-enhanced multiplexed optical lattice clocks~\cite{Zheng2022}, optical clock networks~\cite{Riehle2017}, multi-parameter estimation~\cite{Li2025_multiparam}, distributed sensing techniques~\cite{Malia2022} and protocols free from the gain-bandwidth trade-off~\cite{Qi2025}. We also expect our method of differential measurement to be readily applicable to the search for new physics, such as Lorentz symmetry violation~\cite{Dzuba2016}.

\bibliography{manuscript}

\section{Methods}

\subsection{State dynamics during interrogation}
Working in the rotating frame of reference, where the Hamiltonian of the clock laser drive takes the following form,
\begin{equation}
H=\Delta\ketbra{\text{c}_\downarrow} + \frac{\Omega}{2}\left(\ketbra{\text{c}_\downarrow}{\uparrow}+ \ketbra{\uparrow}{\text{c}_\downarrow}\right), 
\end{equation}
we find the evolution of the initial $\frac{1}{\sqrt{2}}\left(\ket{\downarrow}+\ket{\uparrow}\right)$ state to follow eq. \ref{eq:psi_t}, with:
\begin{equation}
    \begin{split}
        a_\uparrow(\tau)&= \cos\left(\frac{\omega \tau}{2}\right)+i\sin\left(\frac{\omega \tau}{2}\right)\frac{\Delta}{\omega}\\
        a_{\text{c}_\downarrow}(\tau)&=-i\sin\left(\frac{\omega \tau}{2}\right) \frac{\Omega}{\omega},
    \end{split}
\end{equation}
where $\omega= \sqrt{\Omega^2+\Delta^2}$ represents the generalized Rabi frequency. At $\tau=2\pi/\omega$, when the optical qubit undergoes a cyclic evolution, $ a_{\text{c}_\downarrow}=0$ and $ a_{\uparrow}=-1$, and the information about the detuning, $\Delta$, is fully contained in the global phase of the optical qubit (eq. \ref{eqn:aa_phase}).

Note, that the phase is defined modulo $2\pi$, which removes the ambiguity in choosing the encircled area on the Bloch sphere that is identified with twice the global phase. 

In the experiment, we set $\tau=2\pi/\Omega$ and for $\Delta\ne 0$ the evolution is only approximately cyclic. However, the resulting corrections scale as $(\Delta/\Omega)^2$ and affect the ground state superposition along the population axis ($S_z$), whereas the accessed signal is imprinted along the phase axis ($S_y$).

\subsection{Plots of the Wigner quasi-probability distributions}

The plots from Fig. \ref{fig:figure_1} were prepared for illustrative purposes using the tools developed in ref. \cite{Koczor20}.

\subsection{Cavity measurement of the ground state populations}

We extract the $\ket{\uparrow}$ population from the vacuum Rabi splitting of the cavity mode, as in Ref.~\cite{Colombo2022}. Repeating this measurement four times, with RF $\pi$-rotations in between, gives access to both the $\ket{\uparrow}$ and $\ket{\downarrow}$ populations, and allows for the parameter-agnostic read-out of $S_z$, insensitive to the atom loss~\cite{Braverman2019}. The measurement resolution is limited by Raman scattering on the $\ket{\downarrow}\to\ket{{}^3\text{P}_1,m_F=+\tfrac{1}2}$ transition, which introduces noise in $S_z$ during cavity transmission probing~\cite{Braverman2019}. This sets the limit of $8-9$ dB for the maximal amount of resolvable spin squeezing. The quantum amplification protocol can reach higher metrological gains, regardless of this limit \cite{Colombo2022,Davis16}.   

\subsection{Phase control of the clock laser}
Controlling the relative phase of subsequent optical $\pi$-rotations, $\alpha$, involves phase shifts of the RF control signal sent to a double pass acousto-optic modulator (AOM), and introduces technical dead times of $10$ $\mu$s over which the optical qubit evolves freely. This is negligibly short compared to the timescales of the $\pi$-rotations (${\sim}100-400$ $\mu$s).

\subsection{Clock drift cancellation}

The clock laser frequency drifts together with the length of the ultra-low-expansion reference cavity at the constant rate of $0.1$ Hz$/$s. We remove this drift with a double pass AOM fed with the control RF signal that is mixed with a slow ramp at the $-0.1$ Hz$/$s rate. We restart the ramp and reset the clock frequency to the atomic transition every $30$ minutes.

\subsection{Calibration of cavity cooperativity}

\begin{figure}[t!]
    \centering
    \includegraphics[width=0.42\textwidth]{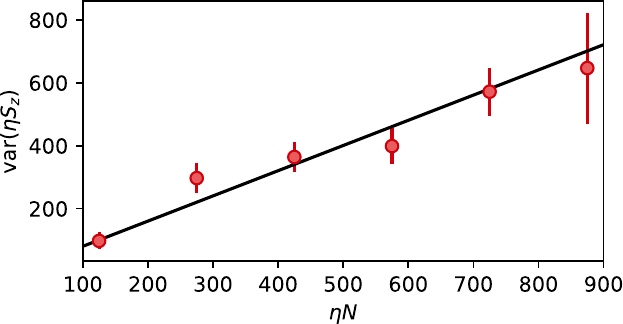}
    \caption{\textbf{Calibration of the effective single-atom cavity cooperativity.} The red points illustrate the measured variances of $\eta S_z$ for binned data sets as a function of the average $\eta N$.
    The black line is a linear fit with slope of $\eta/4$, yielding $\eta=3.2(2)$. Numbers of data points per bin (from left to right): $42$, $104$, $153$, $142$, $163$ and $36$. The size of the error bar of the last point results from a lower number of binned data points, and intrinsically larger $\text{var}(S_z)$ of the CSS for an increased atom number, $N$.
   } 
    \label{fig:figure_SI_coop}
\end{figure}

We calibrate the cavity cooperativity on the $\ket{\uparrow}\to \ket{\text{e}}$ transition, $\eta$, using the method from refs. \cite{Braverman2019,Colombo2022}. This entails measuring the variances in $\eta S_z$ for different $\eta N$, with $\eta=\text{const}$ and $N$ -- the atom number -- varied through the use of different atom loading times. The red points in Fig. \ref{fig:figure_SI_coop} represent the variances of binned datasets, corrected for the measurement resolution effects. The slope of the fitted line, $\eta/4$, reveals $\eta=3.2(2)$. 

\subsection{Differential phase measurement}

To characterize the metrological gain in the quantum-amplified GPS, we employ a differential phase measurement which relies on driving cyclic evolution on both the $\ket{\uparrow}\to\ket{\text{c}_\downarrow}$ and $\ket{\downarrow}\to\ket{\text{c}_\uparrow}$ clock transitions. Immediately after this double interrogation, the atomic state takes the following form:
\begin{equation}
    \ket{\psi}=\tfrac{1}{\sqrt{2}}\left[e^{i\phi_\downarrow} \ket{\downarrow}+e^{i\phi_\uparrow} \ket{\uparrow} \right].
\end{equation}
The imprinted phases, $\phi_\uparrow$ and $\phi_\downarrow$, follow the Eq. \ref{eqn:aa_phase} with transition-specific laser detunings, $\Delta_{\uparrow \to \text{c}_\downarrow}$ and $\Delta_{\downarrow \to \text{c}_\uparrow}$, respectively.  Between the two interrogations, the clock laser frequency is stepped by a precisely calibrated transition frequency difference, $\Delta_{\uparrow \downarrow}=2\pi \times 25.33$ kHz, such that the phases $\phi_\uparrow$ and $\phi_\downarrow$ are mostly dictated by the nearly-common-mode clock laser noise. Since the temporal separation between the two interrogation steps is kept to the shortest allowed by the Rabi frequency of $\Omega=2\pi \times 4.2$ kHz, the phases $\phi_\uparrow$ and $\phi_\downarrow$ are strongly correlated, and the relative phase, $\phi_\uparrow-\phi_\downarrow$, becomes insensitive to the shot-to-shot clock laser noise. Nevertheless, it retains sensitivity to the residual laser noise arising from non-simultaneity of the pulses on distinct transitions, as well as all the effects that would alter the $\ket{\uparrow}\to\ket{\text{c}_\downarrow}$ and  $\ket{\downarrow}\to\ket{\text{c}_\uparrow}$ transition frequencies differently.

\subsection{Relation between differential measurement and the laser-atom comparison}

An actual clock operation entails a read-out of the laser frequency relative to
the (more stable) atomic frequency at a sufficiently long atom-laser interrogation time to precisely measure the laser frequency, as also done in the present work (see Fig. \ref{fig:figure_2}). However, such measurement cannot directly prove operation below the SQL (short of having a second, more stable clock system to compare to, or reverting to near-zero interrogation time) because in the comparison of atomic and laser phases one is simply observing shot-to-shot laser frequency fluctuations.

This is true of
any sub-SQL clock: The atom-laser
clock system uses the atoms to measure the laser frequency, or phase, with a resolution beyond the SQL, but the
directly observed noise is not the atomic sub-SQL noise, but the laser frequency deviation relative to the atoms that
one attempts to correct. The entangled state of the atoms then allows one to correct the laser phase or frequency with a higher resolution than obtainable with an atomic coherent state. 

Our differential measurement constitutes a direct comparison of two stable optical frequencies and provides a workaround to experimentally demonstrate sub-SQL performance of the atomic ensembles, removing the laser frequency noise. The multi-shot statistics from Fig. \ref{fig:figure_4} thus reliably characterize the metrological performance of a single-shot laser-atom frequency comparison.

\subsection{Clock transition cross-talk under the laser drive}
\begin{figure}[t!]
    \centering
    \includegraphics[width=0.42\textwidth]{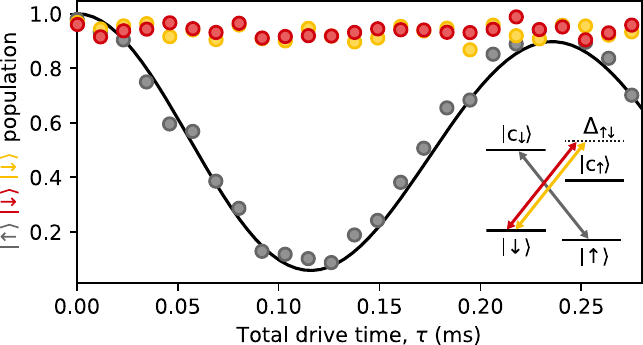}
    \caption{\textbf{Evaluation of the transition cross talk.} The gray points represent resonant Rabi oscillations on an optical clock transition following initialization to the $\ket{\uparrow}$ state. The red and yellow points correspond to Rabi and rotary echo sequences, respectively, following initialization to the $\ket{\downarrow}$ state.
    In all three sequences, the laser frequency was resonant with the $\ket{\uparrow}\to\ket{\text{c}_\downarrow}$ transition.  
   } 
    \label{fig:figure_SI_3Periods}
\end{figure}

We assess the level of crosstalk between the $\ket{\uparrow}\to\ket{\text{c}_\downarrow}$ and $\ket{\downarrow}\to\ket{\text{c}_\uparrow}$ transitions under the laser drive resonant with the former, and conditions close to those of Fig. \ref{fig:figure_4}.
For the measured transition frequency difference, $\Delta_{\uparrow \downarrow}=2\pi \times 25.33$ kHz, and Rabi frequency of $\Omega=2\pi \times 4.2$ kHz, the crosstalk is expected to be negligibly low.

We perform the Rabi and rotary echo sequences following initialization to the $\ket{\uparrow}$ and $\ket{\downarrow}$ states and monitor their subsequent population dynamics (see Fig. \ref{fig:figure_SI_3Periods}). We observe no $\ket{\downarrow}$ population change in both sequences across a single resonant oscillation period. 

In hypothetical scenarios with larger Rabi frequencies, or smaller transition splittings, preventing $\ket{\downarrow}$ population errors can be achieved by tuning the Rabi frequency to satisfy the condition
\begin{equation}
    \Omega=\frac{\Delta_{\uparrow\downarrow}}{\sqrt{n^2-1}}
\end{equation}
for even $n$.

\begin{figure}[t!]
    \centering
    \includegraphics[width=0.45\textwidth]{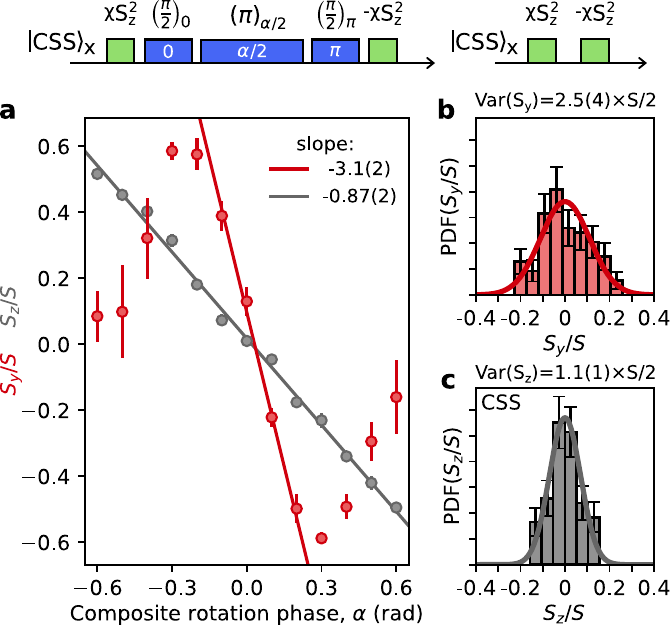}
    \caption{\textbf{Metrological gain in the RF phase encoding scheme.} \textbf{a,} The red (gray) points represent the measured $S_y/S$ ($S_z/S$) as a function of the composite rotation phase, $\alpha$, in the time-reversal (CSS reference) sequence. \textbf{b,} The spin noise statistics in the time-reversal sequence. \textbf{c,} The spin noise statistics in the CSS reference sequence. The `PDF' in panels \textbf{b} and \textbf{c} stands for probability density function, estimated based on more than $100$ measurement outcomes summarized by the histograms. The color coding in the pulse diagrams is consistent with Fig. \ref{fig:figure_1}a-b (Green: entangling light, blue: RF).  
   } 
    \label{fig:figure_SI_benchmarks}
\end{figure}

\subsection{Ground-state benchmarks of metrological gain}

For benchmarking purposes, we study the amplification and spin noise in the absence of the optical phase encoding. The effect of the optically encoded phase, $\alpha$, is mimicked with the composite RF rotation, as outlined in the pulse diagram above Fig. \ref{fig:figure_SI_benchmarks}a.

Using the same parameters of the entangling light as in the experiments behind the Fig. \ref{fig:figure_4}b-d, that is, the same detunings and detected entangling photon numbers ($55(7)$ squeezing and $21(5)$ unsqueezing photons, transmitted through the cavity and collected with ${\sim}8\%$ efficiency), we observe a slope of $-3.1(2)$ in the phase amplification measurement (see the red points in Fig. \ref{fig:figure_SI_benchmarks}a). The signal power is amplified by $11.0(6)$ dB compared to the CSS reference (the gray points in Fig. \ref{fig:figure_SI_benchmarks}a), consistent with the value from the main text within error bars.

Figures \ref{fig:figure_SI_benchmarks}b and \ref{fig:figure_SI_benchmarks}c present the spin noise measured in the time-reversal and CSS reference sequences under the same conditions. The normalized variance in the time-reversal sequence equals $2.5(4)$, and, notably, it is lower than that from 
Fig. \ref{fig:figure_4}c as the probe state is not exposed to the laser noise. This dataset allows us to benchmark the metrological gain with the laser noise subtracted. 

The metrological gain in the ground-state manifold is found as $\mathcal{G}=5.8(9)$ dB.

\subsection{Contrast loss}

\begin{figure}[t!]
    \centering
    \includegraphics[width=0.38\textwidth]{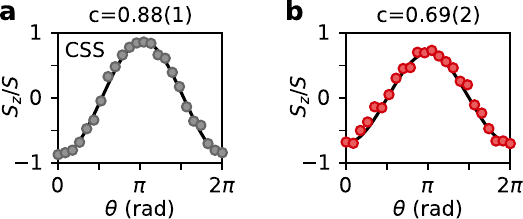}
    \caption{\textbf{Contrast loss from Rayleigh scattering during entangling dynamics.} \textbf{a,} The contrast measured after the CSS reference sequence from Fig. \ref{fig:figure_4}.
    \textbf{b,} The contrast measured after the quantum-amplified sequence from Fig. \ref{fig:figure_4}. 
   } 
    \label{fig:figure_SI_contrast}
\end{figure}

To further verify the coherence preservation in the ground state manifold after quantum-amplified GPS, we perform RF $\pi$-rotations around the $S_y\cos\theta-S_x\sin\theta$ axis, and measure $S_z/S$ as a function of the angle $\theta$. The scan depicted in Fig. \ref{fig:figure_SI_contrast}\textbf{a} corresponds to the CSS reference sequence from Fig. \ref{fig:figure_4}, whereas the one in Fig. \ref{fig:figure_SI_contrast}\textbf{b} to the quantum-amplified sequence from Fig. \ref{fig:figure_4}. The squeezing-unsqueezing dynamics during the quantum-amplified sequence lowers the contrast from $88(1)\%$ to $69(2)\%$ due to single-particle dephasing from Rayleigh scattering of the entangling photons into free space (see \cite{supp}).

\bigskip \noindent \textbf{Acknowledgments:} 
We thank Guoqing Wang, Shai Tsesses and David Spierings for fruitful discussions. This work was supported in part by ONR (grant number N00014-23-1-2577 and DURIP N00014-22-1-2304), the Center for Ultracold Atoms, an NSF funded Frontier Center (grant number PHY-2317134), the NSF Quantum Leap Challenge Institute Award OMA (grant number 2016244), and DARPA (grant number HR00112420357). Support is also acknowledged from the U.S. Department of Energy, Office of Science, National Quantum Information Science Research Centers, Quantum Systems Accelerator.

\bigskip \noindent \textbf{Author contributions} L.Z., Q.L. and G.V. carried out the experimental setup upgrades, with help from E.P.-P., whereas M.R. upgraded the control software. L.Z., Q.L., G.V. and M.R. performed the experiments, simulations and data analysis. Z.L. and S.C. helped with the simulations. V.V. conceived and supervised the experiment.   
L.Z., Q.L. and V.V. wrote the manuscript. All authors discussed the experiment implementation and results and contributed to the manuscript. 

\bigskip \noindent \textbf{Competing interests} The authors declare no competing interests.

\bigskip \noindent \textbf{Data availability} The datasets generated as part of the current study
are available from the corresponding authors upon reasonable request.

\bigskip \noindent \textbf{Code availability} The codes used for the analysis included in the current
study are available from the corresponding authors upon
reasonable request.

\newpage

\end{document}


\title{Supplementary Materials for 'Quantum-amplified Rabi spectroscopy on an optical clock transition'}
\author{Yb clock team and Vladan Vuleti\'c}
\affiliation{MIT-Harvard Center for Ultracold Atoms and Research Laboratory of Electronics, Massachusetts Institute of Technology, Cambridge, Massachusetts 02139, USA }

\section{Noise sensitivity of selected spectroscopic sequences}

\subsection{Phase response functions in time domain}
\begin{figure}
    \centering
    \includegraphics[width=1\textwidth]{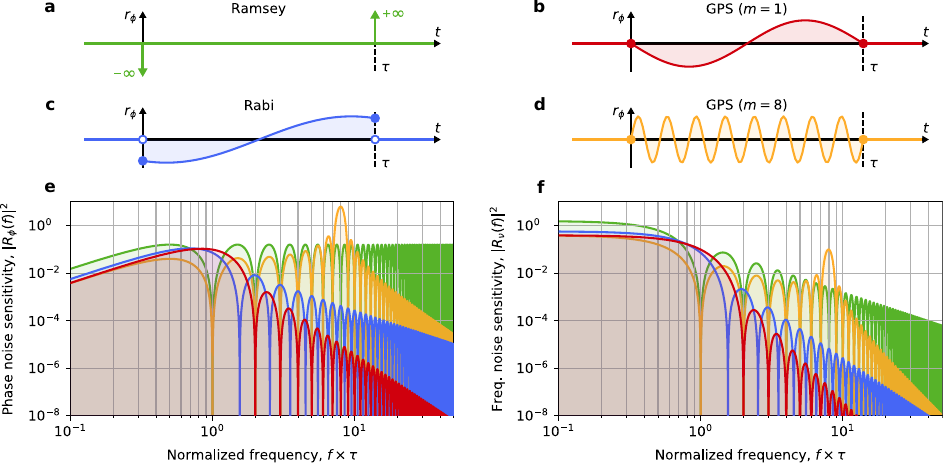}
    \caption{\textbf{Time-domain phase responses and sensitivity functions for selected spectroscopic sequences.} Phase response of \textbf{a,} Ramsey spectroscopy, \textbf{b,} single-cycle GPS, \textbf{c,} Rabi spectroscopy, \textbf{d,} multi-cycle GPS ($m=8$ cycles). \textbf{e,} Phase noise sensitivity functions of considered spectroscopic sequences. \textbf{f,} Frequency noise sensitivity functions of considered spectroscopic sequences. The interrogation time, $\tau$, is the same for all sequences. The color-coding establishes correspondence between panels \textbf{a}-\textbf{d} and panels \textbf{e} and \textbf{f}.  
   } 
    \label{fig:figure_SI_filter}
\end{figure}

\subsubsection{Ramsey Spectroscopy}
Calculation of the phase response function for the Ramsey-type spectroscopic signal is performed most conveniently in the Schr\"odinger picture, starting from: 
\begin{equation}
    \langle \delta S_y\rangle_{\tau}=\frac{1}{2}\sin(\phi(\tau)-\phi(0))=\frac{1}{2}\int_{-\infty}^\infty dt\, \underbrace{[\delta(t-\tau)-\delta(t)]}_{r_\phi(t)}\times \phi(t) +\mathcal{O}(\phi^2),
\end{equation}
where $\delta(t)=\dot{\Theta}(t)$ represents a Dirac function. The time-domain phase response, $r_\phi(t)$, of the Ramsey sequence is plotted in Fig. \ref{fig:figure_SI_filter}a.

\subsubsection{Global Phase Spectroscopy}
To find the GPS phase response we use the Kubo formula \cite{Kubo1957} from linear response theory and investigate the effect of the phase noise on resonant signal:  
\begin{equation}\label{eqn:Kubo}
    \langle \delta S_y\rangle_\tau= -i\int_0^\tau dt\, \bra{\psi_I(0)}[S_{y,I}(\tau),\delta H_I(t)]\ket{\psi_I(0)}.
\end{equation}
The operators involved are expressed in the interaction picture:
\begin{equation}
    \begin{split}
        \delta H_I(t)&=e^{iH_0 t}\delta H(t) e^{-iH_0t}
        \\
         S_{y,I}(t)&=e^{iH_0 t}S_y e^{-iH_0t}
    \end{split}
\end{equation}
with the corresponding Schr\"odinger picture Hamiltonians identified as \cite{Bishof13}:
\begin{equation}\label{eq:bishof_hamiltonian}
\begin{split}
H_0 &= \frac{\Omega}{2}\sigma^{(c)}_x\\
\delta H(t)&=\underbrace{\frac{\Omega}{2}\left(e^{i\phi(t)}-1\right)}_{\epsilon(t)}\sigma^{(c)}_+ + \underbrace{\frac{\Omega}{2}\left(e^{-i\phi(t)}-1\right)}_{\epsilon^*(t)}\sigma^{(c)}_-,
\end{split}
\end{equation}
within the frame co-rotating with the average laser frequency. Here the superscript `$(c)$' refers to operators defined on a clock transition ($\ket{\uparrow}\to\ket{\text{c}_{\downarrow}}$), as opposed to the $S_y$ operator, which is defined in the $\{\ket{\uparrow},\ket{\downarrow}\}$ basis.

We are interested in response after $m$ cyclic evolutions between the ground and clock manifolds, that is, after the time $\tau=m\times 2\pi/\Omega$. Under this condition, we have $S_{y,I}(\tau)=(-1)^mS_y$. Besides, we find:
\begin{equation}\label{eqn:Pert_interaction_pict}
\begin{split}
    \delta H_I(t)=&\left(\epsilon \cos^2\frac{\Omega t}{2} -\epsilon^* \sin^2\frac{\Omega t}{2}\right)\sigma_+^{(c)}\\+&\left(\epsilon^* \cos^2\frac{\Omega t}{2} -\epsilon \sin^2\frac{\Omega t}{2}\right)\sigma_-^{(c)}\\
    +&i\sin\frac{\Omega t}{2}\cos \frac{\Omega t}{2}(\epsilon^*-\epsilon)\sigma_z^{(c)}
\end{split}
\end{equation}
The top two lines of eq. \ref{eqn:Pert_interaction_pict} vanish when substituted in eq. \ref{eqn:Kubo}, leaving us with:
\begin{equation}\label{eqn:Kubo2}
\begin{split}
     \langle \delta S_y\rangle_{\tau}&=-(-1)^m\frac{i}{4}\int_0^\tau dt\,\sin(\Omega t) (\epsilon^*-\epsilon)\\&= -(-1)^m\int_0^\tau dt\,\frac{\Omega}{4}\sin(\Omega t)\times \phi(t) +\mathcal{O}(\phi^2)\\
     &=-(-1)^m\int_{-\infty}^{+\infty}dt\, \underbrace{\frac{\Omega}{4}\sin(\Omega t)[\Theta(t)-\Theta(t-\tau)]}_{r_\phi(t)}\times \phi(t)+\mathcal{O}(\phi^2),
\end{split}
\end{equation}
where $\Theta(t)$ represents a Heaviside step function. Figures \ref{fig:figure_SI_filter}b and \ref{fig:figure_SI_filter}d display the phase responses for GPS with $m=1$ and $m=8$ cyclic evolutions, respectively.

\subsubsection{Rabi Spectroscopy}
To implement a standard Rabi spectroscopy in our setup, one would initialize the atoms in the $\ket{\psi(0)}=\ket{\uparrow}$ state, and measure the $\ket{\uparrow}$ population after $\tau=\frac{\pi}{\Omega}$ of continuous drive with Rabi vector $\boldsymbol{\omega}=(\Omega,0,\Delta)$ for $\Delta\approx0.76\times\Omega$. The Kubo formula from which the effect of laser noise on spectroscopic signal can be calculated is thus: 
\begin{equation}\label{eqn:KuboRabi}
    \langle \delta( \ketbra{\uparrow}{\uparrow})\rangle_\tau= -i\int_0^\tau dt\, \bra{\psi_I(0)}\left[\ketbra{\uparrow}{\uparrow}_I(\tau),\delta H_I(t)\right]\ket{\psi_I(0)},
\end{equation}
and the transformation to the interaction picture uses:
\begin{equation}
    H_0=\frac{1}{2}\boldsymbol{\omega}\cdot \boldsymbol{\sigma},
\end{equation}
with $\boldsymbol{\sigma}=(\sigma_x,\sigma_y,\sigma_z)$ and the superscripts `c' dropped, as the qubit in such a scenario is unambiguously defined on the optical transition.
The Hamiltonian describing the noise ($\delta H(t)=\boldsymbol{\epsilon}(t)\cdot\boldsymbol{\sigma}$) from the bottom line of eq. \ref{eq:bishof_hamiltonian} transforms to:
\begin{equation}\label{eq:RabiKuboHamInt}
    \delta H_I(t)=(\boldsymbol{\epsilon}\cdot\boldsymbol{\sigma})\cos(\omega t)+\underbrace{\boldsymbol{\epsilon}\cdot(\hat{\boldsymbol{\omega}}\times\boldsymbol{\sigma})}_{(\boldsymbol{\epsilon}\times \hat{\boldsymbol{\omega}})\cdot\boldsymbol{\sigma}}\sin(\omega t)+\underbrace{(\boldsymbol{\epsilon}\cdot\hat{\boldsymbol{\omega}})(\hat{\boldsymbol{\omega}}\cdot\boldsymbol{\sigma})(1-\cos\omega t)}_{\mathcal{O}(\phi^2)},
\end{equation}
with $\omega=\sqrt{\Omega^2+\Delta^2}$ and $\hat{\boldsymbol{\omega}}=\boldsymbol{\omega}/\omega$. Since we're interested only in the linear response to the noise, we can ignore the  final term of Eq. \ref{eq:RabiKuboHamInt}.

Similarly:
\begin{equation}\label{eq:rabi_second_operator}
\ketbra{\uparrow}_I(\tau)=\frac{1}{2}\mathds{1}-\frac{1}{2}\left[
(\hat{\mathbf{z}}\cdot\boldsymbol{\sigma})\cos(\omega \tau)+\underbrace{\hat{\mathbf{z}}\cdot(\hat{\boldsymbol{\omega}}\times \boldsymbol{\sigma})}_{(\hat{\mathbf{z}}\times\hat{\boldsymbol{\omega}})\cdot\boldsymbol{\sigma}}\sin(\omega \tau)+(\hat{\mathbf{z}}\cdot\hat{\boldsymbol{\omega}})(\hat{\boldsymbol{\omega}}\cdot\boldsymbol{\sigma})(1-\cos\omega \tau)
\right].
\end{equation}
Because $[\mathbf{a}\cdot\boldsymbol{\sigma},\mathbf{b}\cdot\boldsymbol{\sigma}]=2i(\mathbf{a}\times\mathbf{b})\cdot\boldsymbol{\sigma}$, to find the commutator of the operators from Eq. \ref{eq:RabiKuboHamInt} and Eq. \ref{eq:rabi_second_operator} one needs to calculate six different cross products of classical vectors. However, because of the initial state in Eq. \ref{eqn:KuboRabi}, only the cross products with non-zero $z$-component will contribute to the result (and only their $z$-components will). The relevant $z$-components of all the cross products are calculated below:   
\begin{equation}
 \begin{split}
     [(\hat{\mathbf{z}}\times\hat{\boldsymbol{\omega}} ) \times \boldsymbol{\epsilon}]_z &= -(\boldsymbol{\epsilon}\cdot \hat{\boldsymbol{\omega}})=-\frac{\Omega}{\omega}\epsilon_x(t)\\
     [(\hat{\mathbf{z}}\times\hat{\boldsymbol{\omega}})\times(\boldsymbol{\epsilon}\times\hat{\boldsymbol{\omega}})]_z&=-(\hat{\mathbf{z}}\cdot(\hat{\boldsymbol{\omega}}\times\boldsymbol{\epsilon}))(\hat{\boldsymbol{\omega}}\cdot\hat{\mathbf{z}})=-\frac{\Omega\Delta}{\omega^2}\epsilon_y(t)\\
[\hat{\boldsymbol{\omega}}\times\boldsymbol{\epsilon}]_z&=(\hat{\boldsymbol{\omega}}\times\boldsymbol{\epsilon})\cdot\hat{\mathbf{z}}=\frac{\Omega}{\omega}\epsilon_y(t)
\\
[\hat{\boldsymbol{\omega}} \times(\boldsymbol{\epsilon}\times \hat{\boldsymbol{\omega}} )]_z&=-(\hat{\boldsymbol{\omega}}\cdot\boldsymbol{\epsilon})(\hat{\boldsymbol{\omega}}\cdot\hat{\mathbf{z}})=-\frac{\Omega\Delta }{\omega^2}\epsilon_x(t)
 \end{split}   
\end{equation}
We can further neglect $\epsilon_x(t)$ terms, as they are, to the lowest order, quadratic in $\phi(t)$, while $\epsilon_y(t)\approx-\frac{\Omega}{2}\phi(t)$. This results in:
\begin{equation}\label{eqn:KuboRabifinal}
\begin{split}
    \langle \delta( \ketbra{\uparrow}{\uparrow})\rangle_\tau&= -\int_{0}^\tau dt\, \frac{\Omega^2\Delta}{2\omega^2}\left\{ \sin (\omega t) \sin\left(\pi\frac{\omega}{\Omega}\right)-\cos(\omega t)\left[1-\cos\left(\pi \frac{\omega}{\Omega}\right) \right]\right\}\times \phi(t) +\mathcal{O}(\phi^2)\\
    &= -\int_{-\infty}^\infty dt\, \underbrace{\frac{\Omega^2\Delta}{\omega^2}\sin\left(\frac{\pi\omega}{2\Omega}\right)\sin\left(\omega t-\frac{\pi\omega}{2\Omega}\right)[\Theta(t)-\Theta(t-\tau)]}_{r_\phi(t)}\times \phi(t) +\mathcal{O}(\phi^2)
\end{split}
\end{equation}
and finally we set $\Delta=0.76\times \Omega$, to maximize the DC sensitivity. The phase response function of Rabi sequence in time domain is plotted in Fig. \ref{fig:figure_SI_filter}d.

\subsection{Contribution of the laser noise to spectroscopic signal}

The phase response functions, $r_\phi(t)$, were defined in analogy to ref. \cite{Bishof13}. Keeping with the formalism of ref. \cite{Bishof13}, we find the contribution of the laser noise to a single-shot, single-atom spectroscopic signal:
\begin{equation}
\begin{split}
    \overline{\langle \delta S_y\rangle^2_{\tau}} &= \int_{-\infty}^{\infty} df\, S_\phi(f)|R_\phi(f)|^2 \\
   & =\int_{-\infty}^{\infty} df\, S_\nu(f)\underbrace{\left|\frac{R_\phi(f)}{f}\right|^2}_{|R_\nu(f)|^2} ,
\end{split}
\end{equation}
where bar over the variance represents averaging over noise realizations and $S_\phi(f)$ ($S_\nu(f)$) is a double-sided power spectral density of the laser phase (frequency) noise, whereas $R_\phi(f)$ is a Fourier transform of $r_\phi(t)$. The phase sensitivity function, $R_\phi(f)$, together with the frequency sensitivity function, $R_\nu(f)=R_\phi(f)/f$, thus establish transfer functions between the laser noise spectrum and the noise on spectroscopic signal. The sensitivity functions for considered spectroscopic sequences are plotted in Fig. \ref{fig:figure_SI_filter}e-f.

\subsection{Discussion}
In a spectroscopic task relevant to the operation of an optical lattice clock, one is concerned with a DC response to the laser frequency noise (i.e. in a precise measurement of detuning from the atomic resonance, $\Delta$). Figure \ref{fig:figure_SI_filter}f shows that of all the considered sequences with fixed interrogation time, $\tau$, Ramsey spectroscopy features the highest sensitivity at low frequencies, and is thus a preferred choice for an ideal LO laser. The Rabi spectroscopy and GPS both have a ${\sim}3$ dB lower sensitivity in that range. However, current optical LOs still suffer from AC phase and frequency noise. For high levels of the AC noise, Rabi spectroscopy and GPS are preferred, as they suppress it better than Ramsey spectroscopy (see Fig. \ref{fig:figure_SI_filter}e-f). In particular, the phase noise sensitivity, $|R_\phi(f)|^2$, asymptotes to ${\sim}f^{-2}$ for Rabi spectroscopy and to ${\sim}f^{-4}$ for GPS, while it stays as ${\sim}f^0$ for Ramsey spectroscopy. 
Besides, both the Rabi spectroscopy and the GPS are compatible with extremely low laser intensities, which reduce the systematic effect of the light shift during the excitation \cite{Ludlow15}.

A general understanding of AC-sensitivity hierarchy between the sequences can be built based on the smoothness of time-domain phase responses \cite{Fang_2018}: the Ramsey response is highly singular (Fig. \ref{fig:figure_SI_filter}a), the Rabi response is not singular, but it is discontinuous (Fig. \ref{fig:figure_SI_filter}c), whereas the GPS response is continuous (Fig. \ref{fig:figure_SI_filter}b and Fig. \ref{fig:figure_SI_filter}d). Because of that, the sensitivity of GPS to AC noise is even lower than that of Rabi spectroscopy. We also note that extending the GPS to multiple cycles induces a band-pass sensitivity at a frequency commensurate with the inverse of a cycle time, and while useful for certain spectroscopic tasks~\cite{Bishof13}, it may need to be avoided during the practical clock operation.

\section{Model of the cavity-mediated OAT}

\begin{figure}
    \centering
    \includegraphics[width=0.6\textwidth]{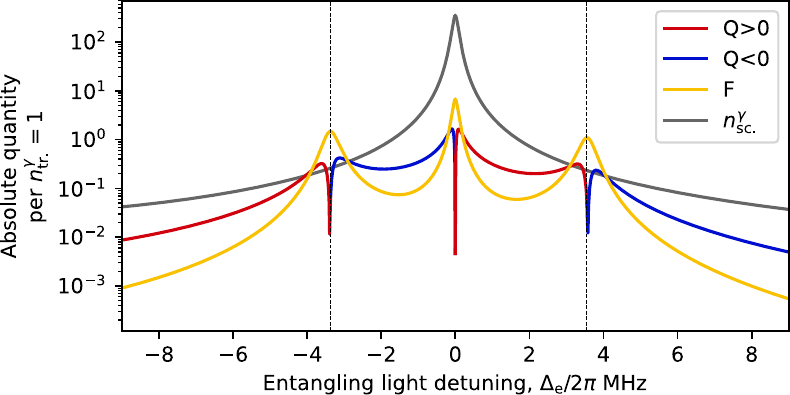}
    \caption{\textbf{Shearing strengths and dephasing levels in a cavity-mediated OAT.} The solid lines display shearing strengths ($Q$), levels of non-unitary broadening ($F$) and Rayleigh scattering rates ($n^\gamma_{\text{sc.}}$) calculated using the microscopic model with parameters from the table \ref{tab:OAT} for a range of entangling light detunings from the $\ket{\uparrow}\to\ket{\text{e}}$ transition ($\Delta_{\text{e}}$). The vertical dashed lines coincide with the vacuum-Rabi-split cavity transmission peaks.    
   } 
    \label{fig:figure_SI_OATmodel}
\end{figure}

\subsection{The microscopics}
We use the results of ref. \cite{ZeyangLi22} to predict the levels of shearing and dephasing during the cavity-mediated OAT. The expressions listed in this section factor in the contributions of the $\ket{\downarrow}\to\ket{{}^3\text{P}_1,\text{m}_{\text{F}}=+\frac{1}{2}}\equiv\ket{\text{e}^\prime}$ transition, detuned from the $\ket{\uparrow}\to\ket{\text{e}}$ transition by $\Delta_{\text{z}}=2\pi \times 22$ MHz. Based on the Clebsh-Gordan coefficients, the cooperativity for the $\ket{\downarrow} \to \ket{\text{e}^\prime}$ transition relates to that for the $\ket{\uparrow} \to \ket{\text{e}}$ transition through $\eta_\downarrow=\eta_\uparrow/3$.  

In line with the ref. \cite{ZeyangLi22}, we introduce non-dimensionalized variables: 

\begin{equation}
    x_a=2\Delta_{\text{e}}/\Gamma,\quad x_c=2(\Delta_{\text{e}}+\delta)/\kappa, \quad b=2\Delta_{\text{z}}/\Gamma
\end{equation}
with $\kappa$ and $\delta$ representing the cavity linewidth and the cavity detuning from the atomic transition, as well as $\Gamma$ denoting the atomic transition linewidth. We also use a shorthand notation for the absorptive and dispersive Lorentzian lineshapes:
\begin{equation}
\mathcal{L}_a(x)=\frac{1}{1+x^2},\quad \mathcal{L}_d(x)=-\frac{x}{1+x^2}.
\end{equation}
Transmission of the lossless, symmetric cavity in the considered scenario is a function of the atomic polarization:
\begin{equation}
    \mathcal{T}_0(S_z) = \frac{1}{[1+(\tfrac{N}{2}+S_z)\eta_\uparrow \mathcal{L}_a(x_a)+(\tfrac{N}{2}-S_z)\eta_\downarrow \mathcal{L}_a(x_a+b)]^2+[x_c+(\tfrac{N}{2}+S_z)\eta_\uparrow\mathcal{L}_d(x_a)+(\tfrac{N}{2}-S_z)\eta_\downarrow\mathcal{L}_d(x_a+b)]^2}.
\end{equation}
The light entering the cavity interacts with the atoms causing (to the lowest order) a constant phase shift and (to the next lowest order) the $S_z$-dependent phase shift, which gives rise to the OAT dynamics. The constant phase shift is refocused with a microwave $\pi$-rotation applied in the middle of the OAT step, and the resulting shearing strength, $Q=N\chi \tau$, is related to the microscopic parameters via:
\begin{equation}\label{eqn:model_Q}
\begin{split}
    Q=&-N\frac{\pi}{\mathcal{F}T_2}\times \mathcal{T}_0(S_z=0)\times[\eta_\uparrow \mathcal{L}_d(x_a)-\eta_\downarrow \mathcal{L}_d(x_a+b)]\\&\times  \bigg\{\left(1+\tfrac{N}{2}\eta_\uparrow \mathcal{L}_a(x_a)+\tfrac{N}{2}\eta_\downarrow \mathcal{L}_a(x_a+b)\right)\times\left( \eta_\uparrow \mathcal{L}_a(x_a)-\eta_\downarrow \mathcal{L}_a(x_a+b)\right)
    \\&+\left(x_c+\tfrac{N}{2}\eta_\uparrow \mathcal{L}_d(x_a)+\tfrac{N}{2}\eta_\downarrow \mathcal{L}_d(x_a+b)\right)\times\left( \eta_\uparrow \mathcal{L}_d(x_a)-\eta_\downarrow \mathcal{L}_d(x_a+b)\right) \bigg\} \times n^\gamma_{\text{tr.}}
\end{split}
\end{equation}
Here, $\mathcal{F}$ stands for the cavity finesse, $T_2$ represents the transmission of the output cavity mirror, and $n^\gamma_{\text{tr.}}$ denotes the number of photons transmitted through the cavity throughout the entire OAT step of duration $\tau$. Note that the transmitted photon number is related to the detected photon number via $n^\gamma_{\text{tr.}}=n^\gamma_{\text{det.}}/\epsilon$, where $\epsilon=0.08$ is the detection efficiency. 

Rayleigh scattering of the photons which enter the cavity causes single-particle dephasing that manifests as an overall contrast loss:
\begin{equation}\label{eqn:c_model}
    c=\exp(-n^\gamma_{\text{sc}.}/N).
\end{equation}
The number of Rayleigh-scattered photons relates to the number of transmitted photons via:
\begin{equation}\label{eq:nsc}
    \frac{n^\gamma_{\text{sc.}}}{n^\gamma_{\text{tr.}}}=N \times\frac{\pi}{\mathcal{F}}\times \frac{1}{T_2}\times [\eta_\uparrow \mathcal{L}_a(x_a)+\eta_\downarrow \mathcal{L}_a(x_a+b)]
\end{equation}
Besides the single-particle dephasing, the atoms experience a collective dephasing due to weak measurements of the atomic state performed by the environment to which the photons exit. This results in a non-unitary broadening of the variance of $S_y$, given by $\text{var}(S_y)\approx\frac{S}{2}(1+F)$, where:
\begin{equation}\label{eq:F_model}
F=N\left(1+\frac{T_1}{T2}R_2+\frac{n^\gamma_{\text{sc.}}}{n^\gamma_{\text{tr.}}}\right)\times \mathcal{T}_0(S_z=0)\times \bigg\{ [\eta_\uparrow \mathcal{L}_a(x_a)-\eta_\downarrow \mathcal{L}_a(x_a+b)]^2+[\eta_\uparrow \mathcal{L}_d(x_a)-\eta_\downarrow \mathcal{L}_d(x_a+b)]^2 \bigg\}\times n^\gamma_{\text{tr.}}
\end{equation}
Here, $T_1$ represents the transmission coefficient of the input mirror, and the $R_2$ ($\approx1$) represents the reflection coefficient of the output mirror. We find our experimental benchmark of $F$ to be consistent with eq. \ref{eq:F_model} within a factor of $2$.

Figure \ref{fig:figure_SI_OATmodel} illustrates $Q$, $F$ and $n^{\gamma}_{\text{sc.}}$ from the equations \ref{eqn:model_Q}, \ref{eq:nsc} and \ref{eq:F_model} plotted for a range of entangling light detunings, $\Delta_{\text{e}}$, and a single transmitted photon (i.e. $n^\gamma_{\text{tr.}}=1$). The underlying model parameters are summarized in the table \ref{tab:OAT}.

\begin{table}[h!]
\begin{tabular}{|c|l|c|}
\hline
Atom number ($N$)                &  & $208$                    \\ \hline
Zeeman splitting of ${}^3\text{P}_1, \, \text{F}=3/2$ ($\Delta_z$) &  & $2\pi \times22 $\,\text{MHz}                 \\ \hline
Atomic linewidth ($\Gamma$)                      &  & $2\pi \times184$\,\text{kHz}                 \\ \hline
Cavity cooperativities ($\eta_\uparrow$ and  $\eta_\downarrow$)             &  & $ 3.2$ and $3.2/3$                \\ \hline
Cavity linewidth  ($\kappa$)               &  & $2\pi \times 796$ kHz \\ \hline
Cavity finesse  ($\mathcal{F}$)               &  & $7540$ \\ \hline
Cavity detuning  ($\delta$)               &  & $2\pi \times 0\,\text{kHz}$  \\
\hline
Transmission of the input cavity mirror ($T_1$)               &  & $30\,\text{ppm}$  \\
\hline
Transmission of the output cavity mirror $+$ loss ($T_2$)               &  & $803\,\text{ppm}$  \\
\hline
Detection efficiency ($\epsilon$)               &  & $8\%$  \\
\hline
\end{tabular}
\caption{The OAT model parameters. \label{tab:OAT}}
\end{table}

\subsection{Spin dynamics}

To combine the above effects into the models of signal and noise, we use the analytical expressions derived in ref. \cite{Schulte20}. 
The amplification factor in SATIN is found as:
\begin{equation}
 M=   -\frac{N-1}{2}c^{(-)}\left(c^{(+)}\right)^2e^{-F^{(-)}/2N}\left(1+e^{-2F^{(+)}/N}\right)\sin\left(Q^{(-)}/N\right)\cos^{N-2}\left(Q^{(-)}/N\right),
\end{equation}
with superscripts `$(+)$' and `$(-)$' marking the quantities relevant to the squeezing and unsqueezing OAT steps, respectively (c.f. eq. \ref{eqn:model_Q}, \ref{eqn:c_model} and \ref{eq:F_model}). 
Note that in the limit of weak dephasing and weak shearing one recovers $M\approx Q$. The expression is also compatible with the strong shearing regime, in which the amplification decays as the Wigner distribution is wrapped around the Bloch sphere ($Q\sim\sqrt{N}$). 

The dynamics away from the linear amplification regime (that is, for encoded phases beyond the dynamic range) are simulated numerically. 

The $S_y$-variance following the two OAT steps is given by:
\begin{equation}
    \text{var}(S_y)=\frac{N}{4}\left\{ 1+\frac{N-1}{2}\left(c^{(+)}c^{(-)}\right)^2\left[1-e^{-2\left(F^{(+)}+F^{(-)}\right)/N}\cos^{N-2}\left(\frac{2\left(Q^{(+)}+Q^{(-)}\right)}{N}\right)\right]\right\},
\end{equation}
and the overall metrological gain follows:
\begin{equation}
\mathcal{G}=M^2\times\frac{N/4}{\text{var}(S_y)}.
\end{equation}

\subsection{Optimal choice of experimental parameters}

We pick the laser detunings, $\Delta_{\text{e}}^{(+,-)}$, and the transmitted photon numbers, $n^{\gamma,(+,-)}_{\text{tr}.}$, for both OAT steps in a way that maximizes the metrological gain, $\mathcal{G}$, subject to  two constraints:
\begin{equation}
\begin{split}
        Q^{(+)}+Q^{(-)}&=0\\
        \frac{\text{d}}{\text{d}N}\left( \frac{Q^{(+)}}{n^{\gamma,(+)}_{\text{in.}}}+\frac{Q^{(-)}}{n^{\gamma,(-)}_{\text{in.}}}\right)&=0.
\end{split}
\end{equation}
The first constraint is that of a time reversal. The second constraint minimizes the sensitivity of the SATIN to atom number fluctuations, for the fixed input photon numbers, $n^{\gamma,(+,-)}_{\text{in.}}$.

\bibliography{SI}